\documentclass[12pt]{article}

\usepackage{graphicx}

\begin{document}

\author{E. Ahmed$^{1,2}$, A. S. Hegazi$^{1}$ and A. S. Elgazzar$^{3}$ \\
$^{1.}$Mathematics Department, Faculty of Science\\
35516 Mansoura, Egypt.\\
E-mail: hegazi@mans.edu.eg\\
$^{2.}$Mathematics Department, Faculty of Science\\
Al-Ain, P. O. Box 17551, UAE.\\
$^{3.}$Mathematics Department, Faculty of Education\\
45111 El-Arish, Egypt.\\
E-mail: elgazzar@mans.edu.eg}
\title{Sato-Crutchfield formulation for some Evolutionary Games}
\date{}
\maketitle

\begin{abstract}
The Sato-Crutchfield equations are studied analytically and numerically. The
Sato-Crutchfield formulation is corresponding to losing memory. Then
Sato-Crutchfield formulation is applied for some different types of games
including hawk-dove, prisoner's dilemma and the battle of the sexes games.
The Sato-Crutchfield formulation is found not to affect the evolutionarily
stable strategy of the ordinary games. But choosing a strategy becomes
purely random independent on the previous experiences, initial conditions,
and the rules of the game itself. Sato-Crutchfield formulation for the
prisoner's dilemma game can be considered as a theoretical explanation for
the existence of cooperation in a population of defectors.\newline
\end{abstract}

\section{Introduction}

Evolutionary game theory [1] is one of the important topics in
Mathematics. Also it has many applications in Biology, Sociology
and Economics. Generally a game consists of a group of players (of
which two players at each game) are interacting with each other
obtaining payoffs. Every player is allowed to choose some
strategies, according to some rules. There are many different
types of games. Each of them has its own rules, strategies,
properties and applications.

Evolutionarily stable strategy (ESS) [1] is an important concept in
population dynamics. Consider a population in which each individual adopts
one of $k$ possible strategies ($I^{j},\;j=1,2,...,k$). A strategy $I^{*}$
is ESS if for all $I^{j}\neq I^{*}$ (yet close to it), then

\begin{enumerate}
\item[(i)]  $A(I^{*},I^{*})>A(I^{j},I^{*})$; or

\item[(ii)]  If $A(I^{*},I^{*})=A(I^{j},I^{*})$, then $%
A(I^{*},I^{j})>A(I^{j},I^{j})$,
\end{enumerate}
where $A$ is the $k\times k$ payoff matrix.

Recently, Sato and Crutchfield [2] have introduced a new formulation that is
corresponding to mistakes or losing memory.

Our aim is to study Sato-Crutchfield formulation both analytically and
numerically. Stability and persistence of the solutions are studied. The
persistence of a dynamical system [3] is defined as follows:\newline
\newline
\textbf{Definition 1:} A dynamical system is persistent if all
cases where all $u_{i}(0)>0$ lead to
\begin{equation}
\lim_{t\rightarrow \infty }\inf u_{i}(t)>0\ \forall i=1,2,...,n.
\end{equation}\\

Also, we study the Sato-Crutchfield formulation for some different
types of two-player games including hawk-dove (HD), prisoner's
dilemma (PD) and the battle of the sexes games [1]. The
Sato-Crutchfield formulation is shown not to affect the ESS of the
ordinary games. But in some limit, choosing a strategy becomes
random independent of the rules of the game and the initial
conditions.

The paper is organized as follow: In section 2, the
Sato-Crutchfield formulation is explained. Both analytical and
numerical results are added. The Sato-Crutchfield equations is
applied to HD game in section 3. Also the effect on ESS is
studied. Section 4 is devoted for studying the Sato-Crutchfield
for PD game and some of its modifications. In section 5, the
Sato-Crutchfield equations are introduced to the battle of the
sexes. Some conclusions are summarized in section 6.

\section{Sato-Crutchfield formulation}

Recently Sato and Crutchfield [2] have studied the dynamics of
learning in multiagent systems, where the agents use reinforcement
learning. They showed that, although the agents are not directly
interacting with each other, a collective game arises between them
through their interaction with environment. Such interactions can
be modelled via replicator type equations.

Consider two agents $u$ and $v$ with $n$ possible actions, let $%
u_{j}(t)\;(v_{j}(t))$ be a measure of the probability that the agent $u(v)$
will take action $j$ at time $t$. Sato and Crutchfield have shown that $%
u_{j}(t),v_{j}(t)$ satisfy the following equations

\begin{equation}
\begin{array}{l}
\frac{\mathrm{d}u_{i}}{\mathrm{d}t}=u_{i}\beta _{1}\left[
(Av)_{i}-uAv\right] +\alpha _{1}u_{i}\sum_{j}u_{j}\ln \left( \frac{u_{j}}{%
u_{i}}\right) , \\
\frac{\mathrm{d}v_{i}}{\mathrm{d}t}=v_{i}\beta _{2}\left[
(Bu)_{i}-vBu\right] +\alpha _{2}v_{i}\sum_{j}v_{j}\ln \left( \frac{v_{j}}{%
v_{i}}\right) ,
\end{array}
\end{equation}
where $A$ and $B$ are payoff (reward) matrices for $u$ and $v$,
respectively, and $\alpha _{1},\alpha _{2},\beta _{1},\beta _{2}$ are
nonnegative constants. If $\alpha _{1}=\alpha _{2}=0$,$\;$then one regains
the standard asymmetric replicator equations [1,4].

We will comment on the case $\alpha _{1}=\alpha _{2}=\alpha >0$. For
simplicity only the symmetric case will be considered $(A=B=\mathrm{diag\;}%
(a_{1},a_{2}),\;u_{1}=v_{1},\;u_{2}=v_{2})$. Rescaling time to set $\beta
_{1}=1$, the system becomes

\begin{equation}
\begin{array}{l}
\frac{\mathrm{d}u_{1}}{\mathrm{d}t}=u_{1}\left(
a_{1}u_{1}-a_{1}u_{1}^{2}-a_{2}u_{2}^{2}\right) +\alpha u_{1}u_{2}\ln \left(
\frac{u_{2}}{u_{1}}\right) , \\
\frac{\mathrm{d}u_{2}}{\mathrm{d}t}=u_{2}\left(
a_{2}u_{2}-a_{1}u_{1}^{2}-a_{2}u_{2}^{2}\right) +\alpha u_{2}u_{1}\ln \left(
\frac{u_{1}}{u_{2}}\right) .
\end{array}
\end{equation}
Restricting to the region $u_{i}\geq 0,\;i=1,2$, then\newline
\newline
\textbf{Corollary 1:} Bounded solutions of the system (3) are persistent.%
\newline
\newline
This means that even bad strategies will be adopted by some agents. To study
this analytically assume $a_{2}>a_{1}>0$. In the case that $\alpha =0$, it
is known that the best strategy is $u_{2}=1,\;u_{1}=0$ so for small $\alpha $%
, we find

\begin{equation}
u_{1}=\exp \left( -\frac{a_{2}}{\alpha }\right) ,\;u_{2}=1-u_{1}.
\end{equation}
The solution (4) is asymptotically stable if

\begin{equation}
\alpha >\exp \left( -\frac{a_{2}}{\alpha }\right) \left(
2a_{1}+4a_{2}\right) .
\end{equation}
The persistence of all strategies can be understood intuitively by realizing
that $\alpha >0$ means memory decay [2]. This implies that some individuals
may not know which is the best strategy hence they choose the bad strategy.
Furthermore as $\alpha $ increases, $u_{1},u_{2}$ become close to each
other, since more agents lose their memory. This agrees with the numerical
simulations shown in Fig. (1). Both $u_{1}$ and $u_{2}$ go to $0.5$ for
large $\alpha $. At this point agents chose their strategies randomly
without any regard to previous experiences.

\section{Sato-Crutchfield formulation for hawk-dove game}

The hawk-dove game [1] is a two-player game, where each player is
allowed to use the hawk strategy (H) or the dove strategy (D). It
has many biological and economic applications. The payoff matrix
is given as follows

\begin{equation}
A=\left[
\begin{array}{cc}
\frac{1}{2}\left( g-c\right) & g \\
0 & \frac{g}{2}
\end{array}
\right] ,
\end{equation}
where $g$ is the value of gain and $c$ is the cost of fight. The
max-min solution of von Neumann et al. [5] is to follow the D
strategy for $0<g<c$. If $0<c\leq g$, then the max-min solution is
to follow the H policy. But this solution is not evolutionary
stable, since a mutant adopting H strategy will gain much, this
will encourage others to adopt the H policy. This will continue
till the fraction of adopting the H policy ($u_{1}$) becomes large
enough to make the payoffs of both strategies equal, thus

\begin{equation}
u_{1}=\frac{g}{c}.
\end{equation}

Applying the Sato-Crutchfield equations to HD game, the replicator
equation becomes

\begin{equation}
\frac{\mathrm{d}u_{1}}{\mathrm{d}t}=\frac{u_{1}}{2}\left[
(g-c)u_{1}u_{2}+gu_{2}^{2}\right] +\alpha u_{1}u_{2}\ln \left( \frac{u_{2}}{%
u_{1}}\right) ,
\end{equation}
where $u_{2}=1-u_{1}$ is the D density. This case is numerically
investigated for various values for $g$ and $c$, see Fig. (2) for $g=1$ and $%
c=4$. The system is updated for $10^{5}$ time steps. The behavior
of $u_{1}$
is studied versus $\alpha $. As shown in Fig. (2), at $\alpha =0$, $%
u_{1}=1/4 $ in agreement with its ordinary value Eq. (7). As
$\alpha $ increases, $u_{1}$ increases rapidly and becomes close
to $0.5$. Therefore, for large $\alpha $, agents choose their
strategies (H or D) randomly with equal probabilities. This
behavior is independent on the initial conditions.

Sato-Crutchfield formulation does not affect the ESS. For example, consider
the hawk(H)-dove(D)-retaliate(R) game, where the payoff matrix is given by

\begin{equation}
A=\left[
\begin{array}{ccc}
\frac{1}{2}(g-c) & g & \frac{1}{2}(g-c) \\
0 & \frac{g}{2} & \frac{g}{2} \\
\frac{1}{2}(g-c) & \frac{g}{2} & \frac{g}{2}
\end{array}
\right] ,\;c>v>0.
\end{equation}
Let $u_{1}(u_{2})$ be the fraction of population adopting H (D)
strategy, and $u_{3}=1-u_{1}-u_{2}$ is the R-density. Then
Sato-Crutchfield equations for the HDR game become
\begin{eqnarray}
\frac{\mathrm{d}u_{1}}{\mathrm{d}t} &=&u_{1}\left[ -\frac{c}{2}+\frac{u_{1}}{%
2\left( c-\frac{g}{2}\right) }+\frac{u_{2}(g+c)}{2}-\frac{u_{1}^{2}c}{2}%
-cu_{1}u_{2}\right]   \nonumber \\
&&+\alpha u_{1}\left[ u_{2}\ln (\frac{u_{2}}{u_{1}})+u_{3}\ln (\frac{u_{3}}{%
u_{1}})\right] , \\
\frac{\mathrm{d}u_{2}}{\mathrm{d}t} &=&u_{2}\left[ u_{1}\left( c-\frac{g}{2}%
\right) -\frac{u_{1}^{2}c}{2}-cu_{1}u_{2}+\alpha \left[ u_{1}\ln (\frac{u_{1}%
}{u_{2}})+u_{3}\ln (\frac{u_{3}}{u_{2}})\right] \right] .  \nonumber
\end{eqnarray}
It is known that for $\alpha =0$, the ESS is R. This means the solution $%
u_{1}=u_{2}=0$ is asymptotically stable. Also, for $\alpha >0$, the ESS is
R, and the solution $(u_{1}=u_{2}=0,u_{3}=1)$ is asymptotically stable.
Beginning from nonzero values for $u_{1},u_{2}$ and $u_{3}$, and for $\alpha
>0$, the system always evolves to a state at which $u_{1}=u_{2}=u_{3}=1/3$.
At this point, selecting a strategy becomes random with equal probabilities.
But beginning from a zero value for the density of a strategy, it will not
grow through the population.

\section{Losing memory and the prisoner's dilemma game}

The existence of cooperation in a group of selfish individuals is an
interesting problem in mathematics and social sciences. One of the models of
this phenomenon is the Prisoner's dilemma (PD) game [1]. In this case the
available strategies are to cooperate (C) or to defect (D). The payoff
matrix $A$ is $\left[ (R,S),(T,P)\right] $, where $T>R>P>S$ and $2R>T+S$.
For simplicity we take $A=$ $\left[ (2,0),(3,1)\right] $.

The standard solution gives the unrealistic conclusion that players should
not cooperate. However cooperation has been shown to be a solution if the
game is modified into iterated PD, or PD on the lattice, or memory PD [6,7].
Recently another modification is given [8], where a third strategy is
introduced so the allowed strategies are C, D and loner (L). In this case,
the payoff matrix is given as
\begin{equation}
A=\left[
\begin{array}{lll}
0 & b & \sigma \\
0 & 1 & \sigma \\
\sigma & \sigma & \sigma
\end{array}
\right]
\end{equation}
where $2>b>1$ and $1>\sigma >0$. We call this game PDL game.

The Sato-Crutchfield equations for the PD game are
\begin{equation}
\frac{\mathrm{d}u_{1}}{\mathrm{d}t}=u_{1}\left(
2u_{1}-2u_{1}^{2}-u_{2}^{2}-3u_{1}u_{2}\right) +\alpha
u_{1}u_{2}\ln \left( \frac{u_{2}}{u_{1}}\right) ,
\end{equation}
where $u_{1}$ is the defectors density and $u_{2}=1-u_{1}$ is the
cooperators density. It is known that for $\alpha =0$, the
asymptotically stable strategy is (D) i.e. $u_{1}=0,\;\,u_{2}=1$.
This has caused the following question to arise: How can
cooperation exist in a population of defectors (selfish
population)? Sato-Crutchfield equations propose a new answer.
Cooperation may arise due to memory loss which is equivalent to
mistakes. But starting from a small number of cooperators in a
defectors population what is the asymptotic ratio of cooperators?
For very small positive $\alpha $ e.g. $O(0.01)$ it is
straightforward to see that the equilibrium solution is
$u_{1}=\exp (-1/\alpha ),\;u_{2}=1-u_{1}$. This solution is
asymptotically stable for very small $\alpha $. But as $\alpha $
increases the solution approaches $u_{1}=u_{2}=0.5\;$(as we have
argued before). Thus in Sato-Crutchfield equations a small number
of cooperators in a population of defectors will increase to reach
$50\%$ of the population
for $\alpha $ not too large, as shown in Fig. (3). This is not the case for $%
\alpha =0$ where it is known that cooperation will eventually tend to zero.
This shows the significant change between Sato-Crutchfield system and the
corresponding replicator one.

Applying Sato-Crutchfield equations to the PDL game, one gets

\begin{equation}
\begin{array}{c}
\frac{1}{u_{1}}\frac{\mathrm{d}u_{1}}{\mathrm{d}t}=u_{2}(1-u_{1})-\sigma
(1-u_{1}-u_{2})(u_{1}+u_{2})-u_{2}^{2} \\
+\alpha \left[ u_{2}\ln (\frac{u_{2}}{u_{1}})+u_{3}\ln (\frac{u_{3}}{u_{1}}%
)\right] , \\
\frac{1}{u_{2}}\frac{\mathrm{d}u_{2}}{\mathrm{d}t}=u_{2}(1-u_{2})-\sigma
(1-u_{1}-u_{2})(u_{1}+u_{2})-bu_{1}u_{2} \\
+\alpha \left[ u_{1}\ln (\frac{u_{1}}{u_{2}})+u_{3}\ln (\frac{u_{3}}{u_{2}}%
)\right] ,
\end{array}
\end{equation}
where $u_{1}$ is the fraction of cooperators, $u_{2}$ is the fraction of
defectors and $u_{3}=1-u_{1}-u_{2}$ is the fraction of loners. For $\alpha =0
$, it is direct to see that $u_{1}=u_{2}=0$ is a neutral equilibrium for the
system (13). For $\alpha >0$, all strategies coexist. Solving the system
(13) numerically for $\alpha =0.1$, we get the fixed point $%
(0.266,0.155,0.579)$. This confirms our conclusion that
Sato-Crutchfield equations is an interesting approach to introduce
cooperation in prisoner's dilemma type games. As $\alpha $
increases, the choice between the three strategies becomes random
with equal probabilities. Therefore $u_{1} $, $u_{2}$ and $u_{3}$
tend to $1/3$.

\section{Sato-Crutchfield equations for the battle of the sexes}

Recently, the war between the sexes was emphasizes as being
important [9]. The battle of the sexes [10] is an asymmetric game
that simulates the conflict between males and females concerning
their respective shares in their parental investment. The female
has two strategies coy or willing (fast) while the male can be
either faithful or philanderer. The male (female) payoff matrix
$A(B)$ is gives as
\[
A=\left[
\begin{array}{cc}
0 & -10 \\
-2 & 0
\end{array}
\right] ,\;\;B=\left[
\begin{array}{cc}
0 & 5 \\
3 & 0
\end{array}
\right] .
\]

Here we are applying the Sato-Crutchfield formulation to the
battle of the sexes. The Sato-Crutchfield equations for the battle
of the sexes are

\begin{equation}
\begin{array}{c}
\frac{\partial u}{\partial t}=u(1-u)(-10+12v)+\alpha _{1}u(1-u)\ln \left(
\frac{1-u}{u}\right) , \\
\frac{\partial v}{\partial t}=v(1-v)(5-8u)+\alpha _{2}v(1-v)\ln \left( \frac{%
1-v}{v}\right) .
\end{array}
\end{equation}
There is a unique internal homogeneous equilibrium solution $E=(5/8,5/6)$
which is stable; but not asymptotically stable. Therefore in the case $%
\alpha _{1}=\alpha _{2}=0$, one gets oscillations. For the case $\alpha
_{1}>0,\;\alpha _{2}>0$, we get\newline
\newline
\textbf{Proposition (1):} The equilibrium solution of the system (14) is $%
u=5/8-(\alpha _{2}/8)\ln 5$,\ $v=5/6+(\alpha _{1}/12)\ln (5/3)$,\ and it is
asymptotically stable $\forall \alpha _{1}>0,\,\alpha _{2}>0$.\newline
\newline
This shows that in this case there is a significant change between
Sato-Crutchfield system and the corresponding replicator one.

\section{Conclusions}

Although the Sato-Crutchfield formulation does not affect the ESS
for a game, there is a significant change between Sato-Crutchfield
formulation and the other formulations of games. Sato-Crutchfield
formulation is corresponding to losing memory. In some limit,
choosing a strategy becomes purely random independent on the
previous experiences, initial conditions, and the rules of the
game itself. This behavior is observed in all the studied systems.
Sato-Crutchfield formulation can be considered as a theoretical
explanation for some aspects in game theory, like the existence of
cooperation in a population of defectors.\\
\\
\section*{Acknowledgments}
We thank D. Stauffer for his helpful comments.

\newpage

\begin{figure}[tbp]
\begin{center}
\includegraphics[angle=-90, width=0.99\textwidth]{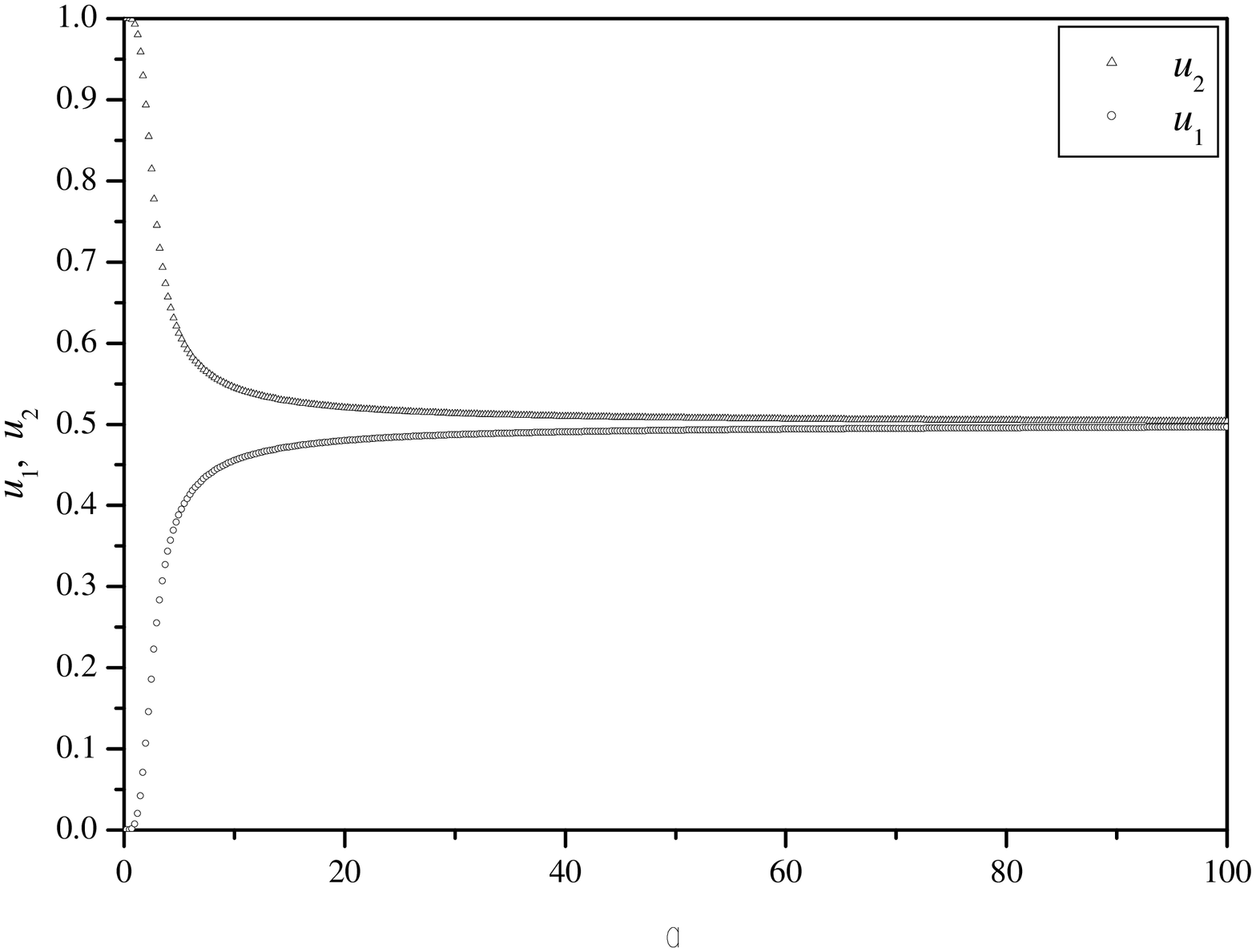}
\end{center}
\caption{For Sato and Crutchfield equations, the densities $u_{1},u_{2}$ are
plotted versus $\alpha$. Both $u_{1}$ and $u_{2}$ go to $0.5$ for large $%
\alpha $. At this point agents chose their strategies randomly.}
\end{figure}

\begin{figure}[tbp]
\begin{center}
\includegraphics[angle=-90, width=0.99\textwidth]{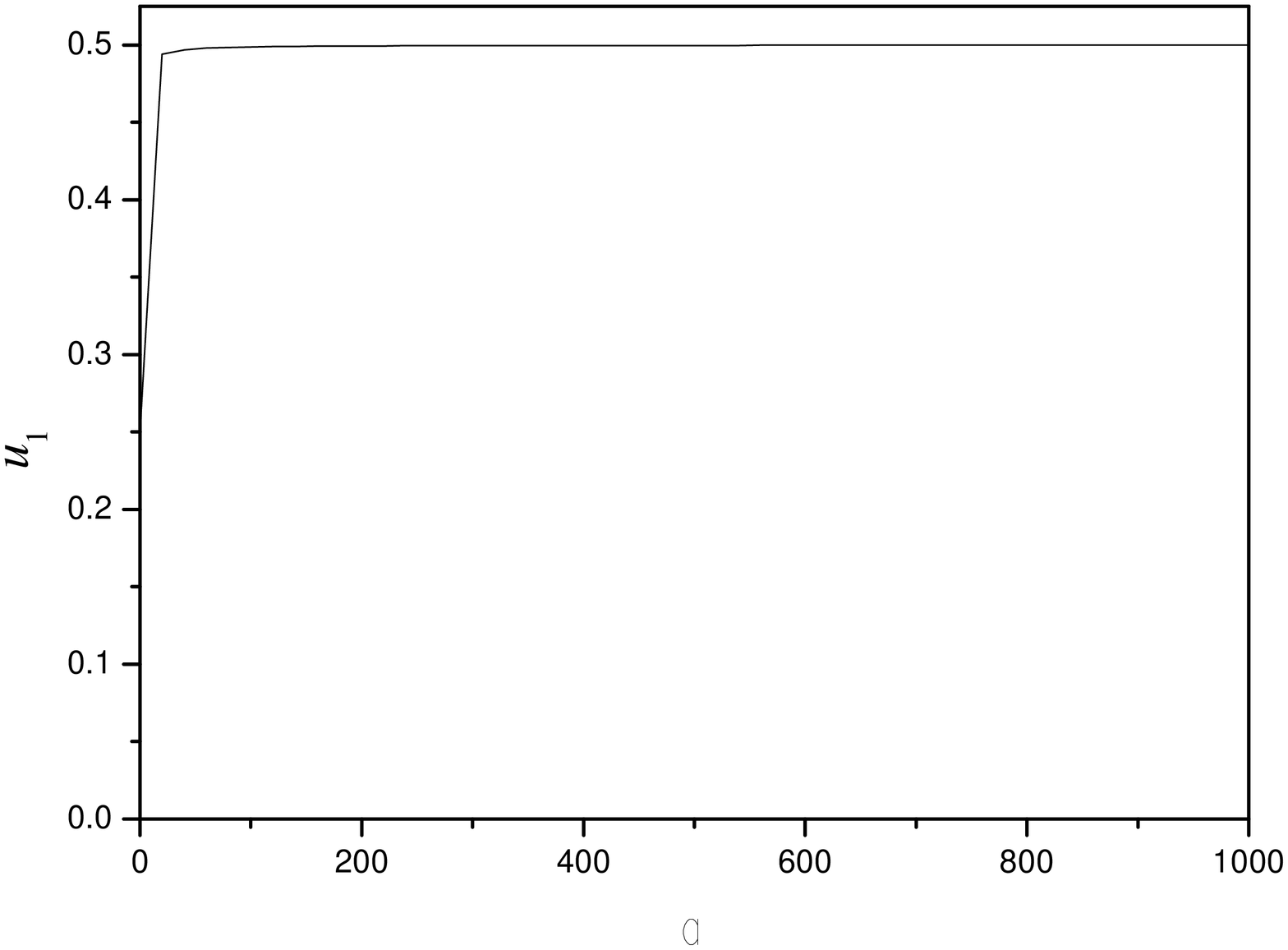}
\end{center}
\caption{In the Sato-Crutchfield formulation for the hawk-dove game, the
density of hawks $u_{1}$ is plotted versus $\alpha$. At $\alpha =0$, $%
u_{1}=1/4 $ in agreement with its ordinary value. As $\alpha$ increases, $%
u_{1}$ increases rapidly and become close to $0.5$.}
\end{figure}

\begin{figure}[tbp]
\begin{center}
\includegraphics[angle=-90, width=0.99\textwidth]{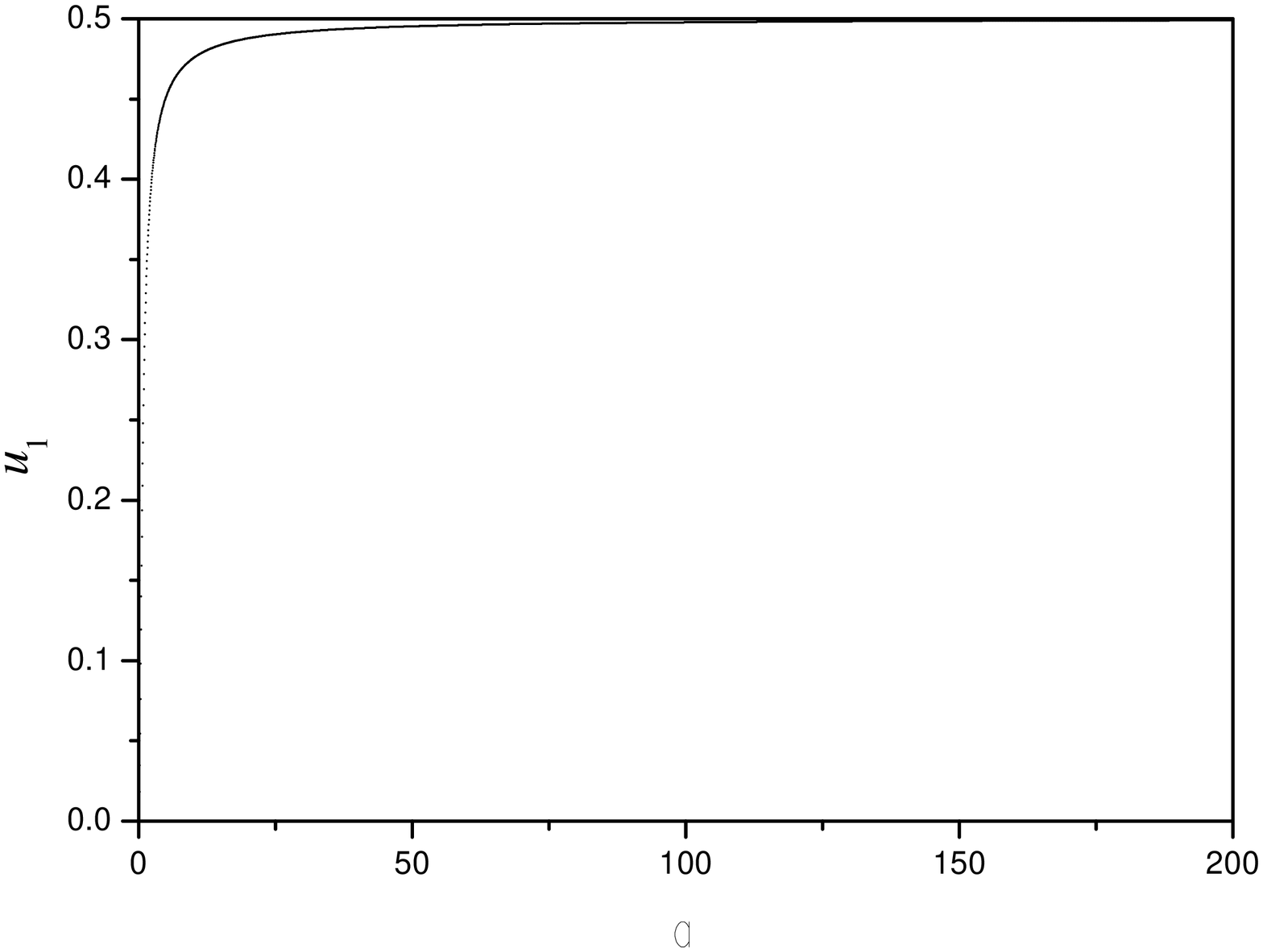}
\end{center}
\caption{Using the Sato-Crutchfield equations for the prisoner's dilemma
game, a small number of cooperators in a population of defectors will
increase to reach 50\% of the population as $\alpha$ increases.}
\end{figure}

\end{document}